# Thermal activation energy of 3D vortex matter in NaFe$_{1-x}$Co$_x$As (*x*=0.01, 0.03 and 0.07) single crystals

W. J. Choi, Y. I. Seo, D. Ahmad and Yong Seung Kwon[*]

Department of Emerging Materials Science, DGIST, Daegu, 42988, Republic of Korea

**Abstract**

We report on the thermally activated flux flow dependency on the doping dependent mixed state in NaFe$_{1-x}$Co$_x$As (*x*=0.01, 0.03, and 0.07) crystals using the magnetoresistivity in the case of *B*//*c*-axis and *B*//*ab*-plane. It was found clearly that irrespective of the doping ratio, magnetoresistivity showed a distinct tail just above the $T_{c,\text{offset}}$ associated with the thermally activated flux flow (TAFF) in our crystals. Furthermore, in TAFF region the temperature dependence of the activation energy follows the relation $U(T,B) = U_0(B)(1 - T/T_c)^q$ with *q*=1.5 in all studied crystals. The magnetic field dependence of the activation energy follows a power law of $U_0(B) \sim B^{-\alpha}$ where the exponent $\alpha$ is changed from a low value to a high value at a crossover field of *B*=~2T, indicating the transition from collective to plastic pinning in the crystals. Finally, it is suggested that the 3D vortex phase is the dominant phase in the low temperature region as compared to the TAFF region in our series samples.

[*] Corresponding author
E-mail *address*: yskwon@dgist.ac.kr (Y. S. Kwon)

**Introduction**

The efforts to find new superconducting materials have recently led to the discovery of various iron-based superconductors [1-4]. Some of the basic similarities with high-$T_c$ cuprates such as their comparable superconducting transition temperature, a two dimensional layered-crystal structure, the dramatic change of superconducting properties by doping and the proximity to a magnetic transition, etc have made iron-based superconductors the center of attention in superconductivity researches [1-4]. Therefore, it is very important aspect to compare these two classes of superconductors to uncover the puzzle of high-$T_c$ superconductivity. However, some important differences are found between these two high-$T_c$ superconducting classes for instance; (1) the parent compound of the iron-based superconductors is not a Mott insulator as in case of the cuprates but a typically bad metal; (2) the iron-based superconductor possesses multiple Fermi surfaces which seems to favor a $s^{\pm}$-wave pairing symmetry rather than a $d$-wave pairing symmetry as in the cuparates; (3) despite the very high values of upper critical field in both classes, less anisotropic nature of iron-based superconductors make them an ideal candidate in terms of technological applications [1-4]. Furthermore, various attempts have also been made in order to determine that the coupling strength between electrons is induced by the Fermi-surface nesting [5] or due to proximity of magnetism [6], however there is no consensus so far.

Among the classes of iron-based superconductors, the 111-type compounds have attracted much attention because even parent compound exhibits superconductivity unlike other classes of iron-based superconductors [7]. For instance, LiFeAs shows bulk superconductivity without doping [8], however parent NaFeAs compound shows ~10% superconducting volume fraction and long-range antiferromagnetic (AFM) ordering along with a structural phase transition [7]. The substitution of Co atom for Fe atomic sites in NaFe$_{1-x}$Co$_x$As, results in bulk superconductivity and suppresses AFM ordering and the superconducting transition temperature increases up to 23 K at 3% Co doping [7]. Recently, it has been reported that the critical current density in optimally doped NaFe$_{1-x}$Co$_x$As (x=0.03) is ~$10^5$ A/ m$^2$ with a distinct peak effect in magnetic hysteresis which is important in terms of physical and applications point of view [9-10]. Therefore, it is crucial to understand the vortex dynamics and the reason for the emergence of peak effect in NaFe$_{1-x}$Co$_x$As. It is well known that in the mixed state, thermal fluctuations directly affect the vortex motion due to thermally activated flux flow (TAFF) in a superconductor. In the case of iron-based superconductors including "111" compounds, relatively much researches have been conducted for the understanding of the vortex dynamics through TAFF analysis [11-16]. There is a lack of systematic studies on the vortex dynamics in 111-type compounds in terms of chemical doping or applied pressure variations. As a result, the vortex dynamics through TAFF analysis still has important issues to deal with such as understanding of the magnetic field dependence of thermally activated energy, the relationship between the thermally activated energy and coherence length, and to understand the anisotropy of the thermally activated energy between in-plane and out-of-plane in these compounds.

In this paper, in order to solve the aforementioned important issues we systematically study the vortex properties in the under doped (x=0.01), optimally doped (x=0.03), and over-doped (x=0.07) $NaFe_{1-x}Co_xAs$ through TAFF resistivity in the mixed state based on electrical resistivity measurements under magnetic fields for $B//c$-axis and $B//ab$-plane. The magnetic field dependence of the activation energy from the TAFF analysis is well understood as a plastic-flux-creep model due to point defects induced by Co-doping.

**Results and Discussion**

Figure 1 (a) shows the single crystalline XRD patterns for $NaFe_{1-x}Co_xAs$ ($x$=0.01, 0.03 and 0.07) single crystals. Only (00$l$) deflection peaks were recognized and showed a full-width-at-half-maximum (FWHM) of ~0.05°, indicating that these single crystals are perfectly c-axis oriented and of high quality. The Co concentration dependence of lattice constant, $c$ of $c$-axis estimated from the XRD patterns is plotted in Fig. 1 (b). The lattice constant $c$ for $x$=0.01 is 7.054 Å, which is consistent with the previous reported values [18]. As shown in Fig. 1(b), the lattice constant decreases linearly with increasing Co concentration indicating the uniform distribution of Co as a dopant in the single crystals, which is again in the agreement with the previous reported results [19].

Figure 2 shows the SEM images and the EDS-spectrum for $NaFe_{1-x}Co_xAs$ single crystals. The EDS mappings indicate that each of the Na, Fe, Co and As elements is homogeneously distributed in the samples which is consistent with XRD results. The molar ratios for each element Na, Fe, Co, As in $NaFe_{1-x}Co_xAs$ with $x$=0.01, 0.03 and 0.07 were estimated to be 1.02:0.99:0.01:1.00, 1.01:0.97:0.03:1.00 and 1.02:0.93:0.07:1.00 with 2% errors, respectively.

Figure 3 shows the temperature dependence of the electrical resistivity under various applied magnetic fields for $B//c$-axis and $B//ab$-plane up to 8 T. In the zero applied magnetic field, the onset (the width) of the superconducting transition is around 19.7 K (4.2 K) for $x$=0.01, 22.8 K (1.7 K) for $x$=0.03, and 20.3 K (1.7 K) for $x$=0.07, respectively. Moreover, the transition width $\Delta T_c$ was estimated as twice the value obtained by the criterion of 10%-50% of the normal-state resistivity. Since, the onset transition temperatures are similar to the values reported previously [19] however the $\Delta T_c$ values in our samples are about 20% less than that of the reported values [19], which is indicative of highly homogeneous samples. The large transition width of $x$=0.01, compared to that of other samples, seems to be due to intrinsic properties caused by the fluctuation of magnetic transition being above $T_c$ [19]. The rounding effect due to the thermodynamic fluctuations of the superconducting cooper pairs was distinctly observed around the onset of the transition temperature in both field-directions in each sample. As the magnetic field increases, the onset transition temperature shifts to the lower temperature side with enhanced rounding effect as shown in Fig. 3. However, in all the samples the shift is more pronounced in the case when $B//c$-axis than for $B//ab$-plane. Furthermore, above the temperature where the electrical resistivity is completely zero, the electrical resistivity showed a distinct tail under magnetic field, which became more evident and extended to lower temperature with increasing magnetic fields. Similar tail has been observed in several layer-structured cuprates and iron-based superconductors and was related with the vortex motion under applied magnetic fields [11-16, 18, 20, 21].

Figure 4 shows the $B_{c2}$-$T$ phase diagram determined by the resistivity data shown in Fig. 3, where $B_{c2}(T)$ corresponds to the temperature where the resistivity drops to the 90% and 50% of the normal-state resistivity in the crystals. As shown in Fig. 4, the obtained $B_{c2}$ is not linear over the entire magnetic field range such that the deviation from linearity is observed above 2 T possibly due to the multiband effect. We estimated the $dB_{c2}/dT$ for our crystals in both field directions using the average gradient above 2 T for the data with 50% criterion and listed it up in Table 1. The obtained values for parameters are similar to the values reported previously [22]. Furthermore, the orbital limiting field at $T$=0 K is given by $B_{c2}^{orb}(0) = 0.693 T_c |dB_{c2}/dT|_{T_c}$ using the Werthamer-Helfand-Hohenberg (WHH) formula within the weak coupling BCS theory [23] and is listed in Table 1. The $B_{c2}^{orb}(0)$ is highest for $x$=0.03 and it is higher for in plane than that of out of plane value. The corresponding Ginzburg-Landau coherence length and the anisotropy ratio defined by $\gamma \equiv \xi_{ab}/\xi_c$ were obtained and are also listed in Table 1. The coherence length and anisotropy ratio have been already reported for $x$=0.021 sample [16]; $\xi_c$ =21.1, $\xi_{ab}$ =34.4 Å and $\gamma$ =1.63, which lie in between the values of our samples for $x$=0.01 and 0.03.

In the mixed state of high $T_c$ superconductor with disorders below $B_{c2}$ they induce barriers for the vortex motion and three different situations can be expected; (1) the energy barrier $U_0$ is lower than the temperature and can be neglected. This state is referred as the unpinned vortex liquid (UVL) state. (2) The energy barrier $U_0$ is higher than the temperature and plays an important role in vortex motion. This corresponds to the thermally activated flux-flow (TAFF) regime. (3) The barrier grows unlimitedly at low critical current density $j$ and the linear resistivity drops to zero. This state is called as the vortex-glass state.

According to the TAFF theory [24], the resistivity in the TAFF and UVL regimes is defined by

$$\rho(T,B) = (2\rho_c U/T) \exp(-U/k_B T) = \rho_{0f} \exp(-U/T), \quad (1)$$

where $\rho_{0f}$ is the temperature-independent constant, $U$ is the thermal activation energy and $k_B$ is Boltzmann's constant. To obtain the activation energy $U(T, B)$ we consider the temperature dependence of the derivative $D = \partial(\ln \rho)/\partial(T^{-1})$ at various applied magnetic fields and both field directions.

The representative results are depicted in the upper panel of Fig. 5. In the Arrhenius relation of $U(T,B) = U_0(B)(1 - T/T_c)$, where $U_0$ is the apparent activation energy and usually plot shows a distinct plateau in the TAFF region. As shown in the figures, however, the plateau is not seen but rather increases as temperature decreases. More specifically, the $D$-value is almost zero above $T_K$, which is characteristic of the UVL state. In $T^* < T < T_K$ expected as TAFF region the $D$-value increases relatively slowly, which indicates that studied samples have a non-linear activated energy of the form of $U(T,B) = U_0(B)(1 - T/T_c)^q$ [18, 25]. In $T<T^*$ the $U$-value increases faster and then diverges, which signifies $U(T, B)$ as entering into the vortex-glass critical-state region as previously seen in high-$T_c$ cuprates and FeAs superconductors [11, 25, 26]. Under the non-linear activation energy in the TAFF regime we can derive the equation of the form

$$\ln \rho = \ln 2\rho_c U_0 + q \ln(1 - T/T_c) - \ln T - U_0(1 - T/T_c)^q/T. \qquad (2)$$

To obtain the activation energy we plotted the $\ln \rho$ versus $1/T$ data for both magnetic field directions of $B//c$-axis and $B//ab$-plane at $x=0.01$, 0.03 and 0.07 in Fig. 6 and tried to fit the data using eq. (2). We found the value of the exponent $q=1.5$ in both field directions for all the samples and the value of $T_c$ as 16.8, 20.9 and 16.8 K for $x=0.01$, 0.03 and 0.07, respectively. The $\rho_c$ almost linearly increased with applied magnetic fields as shown in Fig. 7. According to the condensation model, $q=1.5$ is expected in the case of high $T_c$ superconductors which show 3D vortex behavior, whereas $q=2$ represents 2D vortex behavior [27-29]. The magnetic field dependence of $U_0$ evaluated from this fitting is plotted in Fig 8 (a) and (b) for $B//c$-axis and $B//ab$-plane for $x=0.01$, 0.03 and 0.07, respectively. The characteristics of the determined $U_0$-value are as follows:

(1) At $x=0.01$ (underdoped sample), $U_0$ decreases from 752 K (1213 K) at $B=0.5$ T to 207 K (354 K) at $B=8$ T with increasing applied magnetic field for $B//c$-axis ($B//ab$-plane). The field dependence of $U_0$ follows a power law relation $[U_0(B) \sim B^{-\alpha}]$ for both field-directions. The exponent α has a small value of 0.27 up to $B=\sim 2$ T and a large value $\alpha = 0.73$ over ~2 T; for $B//c$-axis, for $B//ab$-plane, $\alpha = 0.29$ when $B < \sim 2\,T$ and $\alpha = 0.70$ when $B > \sim 2\,T$.

(2) At $x=0.03$ (optimally doped sample), $U_0$ decreases from 8400 K (11250 K) at $B=0.5$ T to 724 K (3216 K) at $B=8$ T with increasing magnetic field for $B//c$-axis ($B//ab$-plane). These values are significantly increased for both field directions as compared with the underdoped sample. The $U_0$-value is similar to the values reported for SmFeAsO$_{0.9}$F$_{0.1}$($T_c$=54 K) [25] and SmFeAsO$_{0.85}$ ($T_c$=50 K) [11]. The power law relation was also observed in the field dependence of the activation energy in case of the optimally doped sample. The exponent significantly increases for $B//c$-axis but remains almost unchanged for $B//ab$-plane; $\alpha = 0.73$ when $B < \sim 2\,T$ and $\alpha = 1.16$ when $B > \sim 2\,T$ for B//c-axis and $\alpha = 0.28$ when $B < \sim 2\,T$ and

$\alpha = 0.69$ when $B > \sim 2\,T$ for B//ab-plane.

(3) At $x$=0.07 (overdoped sample), $U_0$ decreases from 2470 K (4360 K) at $B$=0.5 T to 330 K (1277 K) at $B$=8 T with increasing applied magnetic field for $B$//$c$-axis ($B$//$ab$-plane). These values are between those of the underdoped and optimally doped samples. The field dependence of $U_0$ in the underdoped sample also follows a power law relation for both directions. Furthermore, the exponents are between those of the underdoped and optimally doped samples for $B$//$c$-axis but is almost unchanged for $B$//$ab$-plane; $\alpha = 0.52$ when $B < \sim 2\,T$ and $\alpha = 1.05$ when $B > \sim 2\,T$ for $B$//c-axis and $\alpha = 0.29$ when $B < \sim 2\,T$ and $\alpha = 0.67$ when $B > \sim 2\,T$ for $B$//ab-plane.

As mentioned earlier, the activation energy of NaFe$_{1-x}$Co$_x$As follows a power law relation with different value of exponent α near $B$=~2 T; in the low magnetic fields α is small, whereas in high magnetic fields α is large similar to iron-based superconductors [12, 13, 30, 31]. Nevertheless, the field dependence of $U_0$ following a power law has not been well understood so far. Since, as the magnetic field increases, the reduction of $U_0$ in the form of $U_0(B) \sim B^{-\alpha}$ has been described as a result collective elastic creep [32]. However, according to the collective elastic creep theory the activation energy should increase inversely [33-35].

Very recently, it has been reported that the collective pinning which dominates below $B$=~2 T crossover to the plastic pinning for $B > \sim 2$ T observed from magnetic properties near $T_c$/2 in NaFe$_{1-x}$Co$_x$As ($x$=0.01, 0.03, 0.05 and 0.07) for $B$//c-axis [9]. The crossover field shifted to low magnetic field as the temperature increased. As a result of this crossover, the secondary peak was observed in magnetic hysteresis in the optimally doped and overdoped samples but it was absent in the underdoped sample [9]. The peak seems to appear as a crossover from the collective pinning to the plastic pinning, but it was not well-understood about the absence of the seconday peak effect in the underdoped sample.

From this point of view, it is reasonable to assume that the field dependence of $U_0$ is governed by both the collective pinning and plastic pinning. According to the plastic-flux-creep theory [36-38], the vortices are plastically deformed and entangled by the weak pinning with point defects, as a result the field dependence of activation energy follows the form $U_0 \sim B^{-0.5}$. In high magnetic fields, the entangled vortices are cut and disconnected due to the faster motion of vortices relative to each other so the activation energy can be described in the form of $U_0 \sim B^{-0.7}$ [26, 36]. The faster reduction in $U_0$ with magnetic fields was suggested to be due to the entangled vortex liquid behavior in a region with strong pinning by point defects [26].

Furthermore, by taking into account the magnetic measurement results of NaFe$_{1-x}$Co$_x$As ($x$=0.01, 0.03 and 0.07) [9], the crossover field in the temperature range of the TAFF analysis may shift to lower than 2 T. Therefore, it is reasonable to assume that the coexistence of the plastic pinning and collective pinning coexist below 2 T in this temperature range. At $x$=0.01 for $B$//$c$-axis, in low magnetic field range the exponent is smaller than 0.5, the value expected for the plastic pinning as mentioned above, which is well understood by the coexistence of the collective weak pinning and plastic weak pinning with point defects. However, in the high magnetic fields the exponent is equal to 0.7 in accordance with the plastic weak pinning theory in high magnetic field range. At $x$=0.03 and 0.07 in $B//c$-axis, the both exponents for low and high magnetic fields exceed the expected values which correspond to plastic weak pinning theory as mentioned above, which may indicate the presence of the plastic strong pinning. However, the reason that the exponent in low magnetic field is smaller than that in higher magnetic field may be due to the coexistence of collective pinning and plastic pinning. In case of $x$=0.07, the plastic pinning becomes more prominent because of lower crossover field [9] and the exponent in low field region is predicted to be larger, but it is smaller in our result. This indicates that the strength of the pinning for $x$=0.07 is weaker than that for $x$=0.03.

In case of $B//ab$, three samples show similar magnetic dependence of $U_0$. In case of $B//c$-axis, we considered that point defects in NaFe$_{1-x}$Co$_x$As play an important role in the emergence of plastic pinning which lead to entangled vortex liquid behavior. The entanglement of vortex lines with increasing magnetic field reduces the correlation length along vortex lines which results in to the reduction of the activation energy. It is expected that the smaller the degree of entanglement of the vortex line, weaker the decreasing trend of the activation energy with increasing magnetic field. In NaFe$_{1-x}$Co$_x$As superconductor, Co ions are thought to act as mainly point-pinning centers. Co-ions concentration at $x$=0.03 and 0.07 increases along the c-axis but is almost unchanged in the ab-plane when compared with the sample with $x$=0.01, which is determined from the variation of lattice constants. Since, with Co doping, the lattice constant of $a$ is almost unchanged but the lattice constant of $c$ is linearly decreased with increasing Co [19]. The correlation length of vortex lines for $B//ab$-plane is expected to be approximately the same in the three samples and show similar magnetic field dependency regardless of whether the pinning is weak or strong.

Analysis of the electrical resistivity below the TAFF temperature region can confirm the presence of the vortex-glass state in NaFe$_{1-x}$Co$_x$As superconductor. The vortex glass theory predicts the temperature dependence of the linear resistivity through the relation $\rho(T) \propto (T - T_g)^s$ in the vortex-glass critical region just above the glass transition temperature $T_g$, where $s$ is the critical exponent [39]. As shown in Fig. 9, the linear resistivity below $T^*$ is well described by this equation and a straight line is shown in the critical region of the glass transition in a plot of $[\partial \ln \rho / \partial T]^{-1}$ versus $T$. The critical exponent $s$ is plotted in insets of Fig. 9. The $s$-values for different samples are almost independent of magnetic field in both magnetic field directions and are larger than $s$=2.7, which is the lower limit of $s$ predicted by the 3D vortex-glass picture [32]. The 3D vortices in vortex liquid, which are determined from the $q$-values discussed above, are frozen in to the 3D vortex glass with decreasing temperature, in agreement with the vortex glass theory [39]. The vortex glass transition temperature $T_g$ determined by the extrapolation is shown in Fig. 4. The vortex glass critical temperature $T^*$ drawn in Fig. 4 is defined as the temperature which deviates from the linear electrical resistivity in vortex glass state and is approximately equal to the temperature on the lower side deviating from the TAFF fitting (Fig. 5) .

As shown in Fig. 4, the vortex matter phase diagram remarkably depends on the Co-doping concentration. The characteristic fields such as $B_g(T)$, $B^*(T)$ and $B_k(T)$ shift to low temperature region in a sample with low activation energy and low $T_{c2}$. The $B_k(T)$ is the characteristic magnetic field that divides the vortex liquid phase in to a pinned liquid phase and a unpinned liquid phase and is defined as the temperature on the higher side deviating from the TAFF fitting (Fig. 5) . This crossover field of $B{\sim}2$ T exists in the magnetic field dependence of the activation energy, which is plotted in the pinned liquid region as $B_{cr}$. For $x=0.01$, the vortex liquid is in a state of coexistence of the weakly collective pinning and the weakly plastic pinning below $B_{cr}$ and it is in a state of weakly plastic pinning above $B_{cr}$. For $x=0.03$ and 0.07, the collective pinning and the plastic pinning coexist under the strong pinning limit in $B<B_{cr}$ but the plastic pinning only exists in $B>B_{cr}$. From this, it is likely that the peak effect observed in the magnetic hysteresis for $x=0.03$ and 0.07 occurs when crossing from a collective pinning to a plastic pinning in a strong pinning.

**Conclusions**

We investigated the thermally activated flux flow (TAFF) using the magnetoresistivity in single crystals NaFe$_{1-x}$Co$_x$As ($x=0.01$, 0.03, and 0.07) for $B//c$-axis and $B//ab$-plane. Results showed that just above the temperature where the electrical resistivity is completely zero, the electrical resistivity shows a distinct tail as a result of thermally activated flux flow which becomes more evident with increasing magnetic fields. This resistivity behavior due to TAFF is well understood by considering the nonlinear temperature relationship of the activation energy. The magnetic field dependence of the estimated activation energy follows a power law of $U_0(B){\sim}B^{-\alpha}$ but the exponent increases above the applied magnetic field of $B={\sim}2T$ subjected to vortex phase transition. We have found that that in NaFe$_{1-x}$Co$_x$As, crystals, point defects formed as a result of Co doping has played an important role in the vortex pinning properties of our samples. Furthermore, it is suggested that in the underdoped ($x=0.01$) sample the weak pinning mechanism plays a dominant role in TAFF region whereas in the case of optimally doped ($x=0.03$) and overdoped (0.07) samples the strong pinning mechanism seems to be effective.

**Methods**

Single crystals of NaFe$_{1-x}$Co$_x$As (x=0.01, 0.03, and 0.07) were grown by the Bridgman method in a vertical vacuum furnace with temperature stability of ±0.5°C or less [17]. Before growing single crystals, FeAs precursor was prepared by reacting Fe-powder and As-lump at 1050°C for 120 h. After that, Na: Fe: Co: As were weighed in a molar ratio and put it in a BN crucible, and then put all in a molybdenum (Mo) crucible. This all procedure was carried out in a glove box having purified argon (Ar) gas atmosphere with the level of $H_2O$ and $O_2$ contents is 0.1 ppm or less. The Mo crucible was arc welded in Ar gas atmosphere to avoid the escape of highly volatile As and Na elements. In the final stage, the sealed Mo crucible was heated up to 1400 °C at a rate of 60 °C/h and kept for 72 hrs in a vertical vacuum electric furnace composed of tungsten mesh, after that the crucible was slowly moved to the bottom of the heater at a rate 1.7 mm/h. After the completion of the heat treatment, a number of single crystals of size 5 x 5 x 1mm$^3$ were obtained.

In order to analyze the crystalline structure and lattice constant, x-ray diffraction (XRD) for a cleaved surface of single crystals was performed with a Cu-K$_\alpha$ radiation source. The detailed analysis of the compositions of the single crystals was performed using an energy-dispersive x-ray spectroscopy (EDS) and mapping method. The EDS mappings were measured by a Hitachi S-4800 scanning electron microscope with an energy dispersive X-ray analyzer (Bruker QUANTAX). The accelerating voltage and the applied current were 20 kV and 10 µA. The EDS mappings were performed for each of the Na, Fe, Co and As elements.

The electrical resistivity measurements of series single crystals were carried out by using a 9 T Physical Property Measurement System (PPMS, Quantum Design, Inc.). Since, the samples are highly sensitive to air and moisture, therefore all the work such as sample cutting and lead-wires connection was performed in the glove box mentioned above. The transversal resistivity was measured down to 2 K using a standard four-probe method at a current density of ~1 A/cm$^2$. The applied magnetic fields of B = 0, 0.5, 0.75, 1, 1.5, 2, 3, 4, 5, 6, 7 and 8 T were used for both field directions as *B*//*c*-axis and *B*//*ab*-plane for measuring the magnetoresistivity. The samples used in this study were taken from the same batches that were used in our recent study on magnetic properties and specific heat results [9].

**Acknowledgements (not compulsory)**

This work was supported by the NRF grant funded by the Ministry of Science, ICT and Future Planning (2015M2B2A9028507, 2016R1A2B4012672 and 2012K1A4A3053565).

**Author contributions statement**

W.J.C., Y.I.S. and D.A. performed magnetoresistivity experiment. W.J.C. prepared and characterized single crystalline samples. Y.S.K. and W.J.C. wrote the manuscript. All authors discussed the results and reviewed the manuscript.

**Additional information**

**Competing financial interests:** The authors declare no competing financial interests.

**Figure and Table captions**

Table 1. Values of the parameters determined from the resistivity data with 50% criterion for $x$=0.01, 0.03, and 0.07 for NaFe$_{1-x}$Co$_x$As.

Fig. 1. (a) XRD patterns for NaFe$_{1-x}$Co$_x$As single crystals. (b) The lattice constant of $c$-axis plotted as a function of the Co concentration $x$. The lattice constant of $c$-axis for $x$=0 is cited from the previous report [18]. Inset in (a) is the rocking curve for the (002) deflection of the single crystal for $x$=0.01.

Fig. 2. EDS analysis for NaFe$_{1-x}$Co$_x$As single crystals; (A) $x$=0.01, (B) $x$=0.03, and (C) $x$=0.07. (a) SEM image of the single crystal, and (b) the corresponding EDS mapping image for all elements and each separate element (c) Na (d) Fe (e) Co and (f) As. (G) EDS spectrum of the NaFe$_{1-x}$Co$_x$As single crystals.

Fig. 3. Temperature dependence of the electrical resistivity of NaFe$_{1-x}$Co$_x$As single crystals under various applied magnetic fields; (a) $x$=0.01 for $B$//$c$-axis, (b) $x$=0.01 for $B$//$ab$-plane, (c) $x$=0.03 for $B$//$c$-axis, (d) $x$=0.03 for $B$//$ab$-plane, (e) $x$=0.07 for $B$//$c$-axis, and (f) $x$=0.07 for $B$//$ab$-plane.

Fig. 4. Static phase diagram of NaFe$_{1-x}$Co$_x$As. $B_{c2}^{90\%}$ and $B_{c2}^{50\%}$ are the upper critical fields estimated from 90% and 50% of the normal state resistivity, respectively. The characteristic fields $B_g(T)$, $B^*(T)$ and $B_K(T)$ are determined from the temperature corresponding to the vortex glass-to-liquid transition, the upper-field limit of the critical region associated with vortex glass-to-liquid phase transition and the temperature that separates the vortex liquid phase from the pinned vortex liquid phase, respectively. The crossover field $B_{cr}(T)$ is determined from the field dependence of the activation energy.

Fig. 5. Plots for $-(d (\ln\rho)/d(1/T))$ vs. $T$ (highest panel), $\ln\rho$ vs. $T$ (middle panel) and $[d(\ln\rho)/dT]^{-1}$ vs. $T$ (lowest panel) at $x$=0.01 (a), 0.03 (b) and 0.07 (c).

Fig. 6. The ln ρ($T$) versus 1/$T$ curves and the curves fitted by the TAFF model by considering the temperature-dependent prefactor and nonlinear relation of $U(T,B) = U_0(B)(1 - T/T_c)^q$ at $x$=0.01 (highest panel), 0.03 (middle panel), and $B$//$c$-axis (lowest panel) under $B$//$c$-axis and

*B*//*ab*-plane.

Fig. 7. Plot on the field dependence of parameter $\rho_c(B)$ determined in TAFF analysis.

Fig. 8. The thermally activated energy $U_0(B)$ determined by the TAFF model with nonlinear relation of $U(T,B) = U_0(B)(1 - T/T_c)^q$ for *B*//*c*-axis (a) and *B*//*ab*-plane (b). The magnetic field dependence of the activation energy normalized by the $U_0$ at *B*=0.5 T for *B*//*c*-axis (c) and *B*//*ab*-plane (d).

Fig. 9. Inverse logarithmic derivative of resistivity for different fields at *x*=0.01, 0.03 and 0.05. The solid red lines represent fits to the vortex-glass theory using the relation of $[d(\ln\rho)/dT]^{-1} = (1/s)(T - T_g)$. Insets represent the magnetic field dependence of *s*(*B*).

Table 1. Values of the parameters determined from the resistivity data with 50% criterion for $x$=0.01, 0.03, and 0.07 for NaFe$_{1-x}$Co$_x$As.

| | $|dB_{c2}/dT|_c$ (T/K) | $|dB_{c2}/dT|_{ab}$ (T/K) | $T_c$ (K) | $B_{c2,c}$ (T) | $B_{c2,ab}$ (T) | $\xi_c$ (Å) | $\xi_{ab}$ (Å) | $\gamma$ |
|---|---|---|---|---|---|---|---|---|
| $x$=0.01 | 2.15 | 3.41 | 17.2 | 25.5 | 40.5 | 22.6 | 35.9 | 1.6 |
| $x$=0.03 | 2.57 | 7.87 | 21.3 | 37.8 | 115.7 | 9.6 | 29.5 | 3.1 |
| $x$=0.07 | 2.68 | 5.98 | 17.6 | 32.5 | 72.6 | 14.3 | 31.8 | 2.2 |

Fig.1

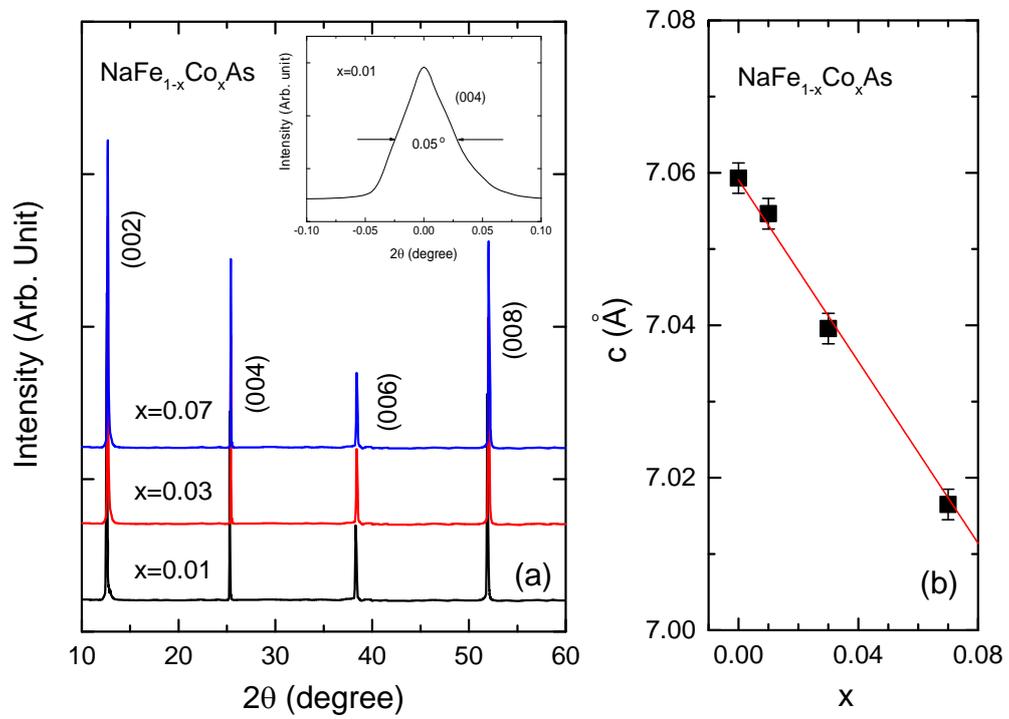

Fig. 2

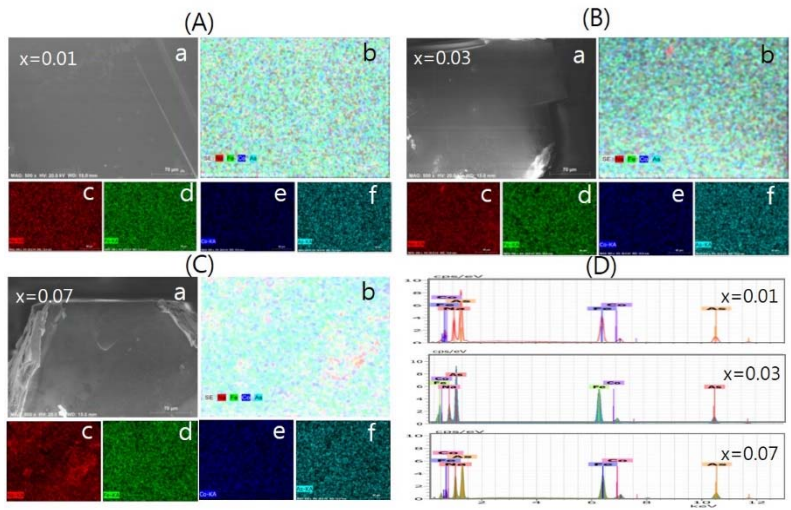

Fig. 3

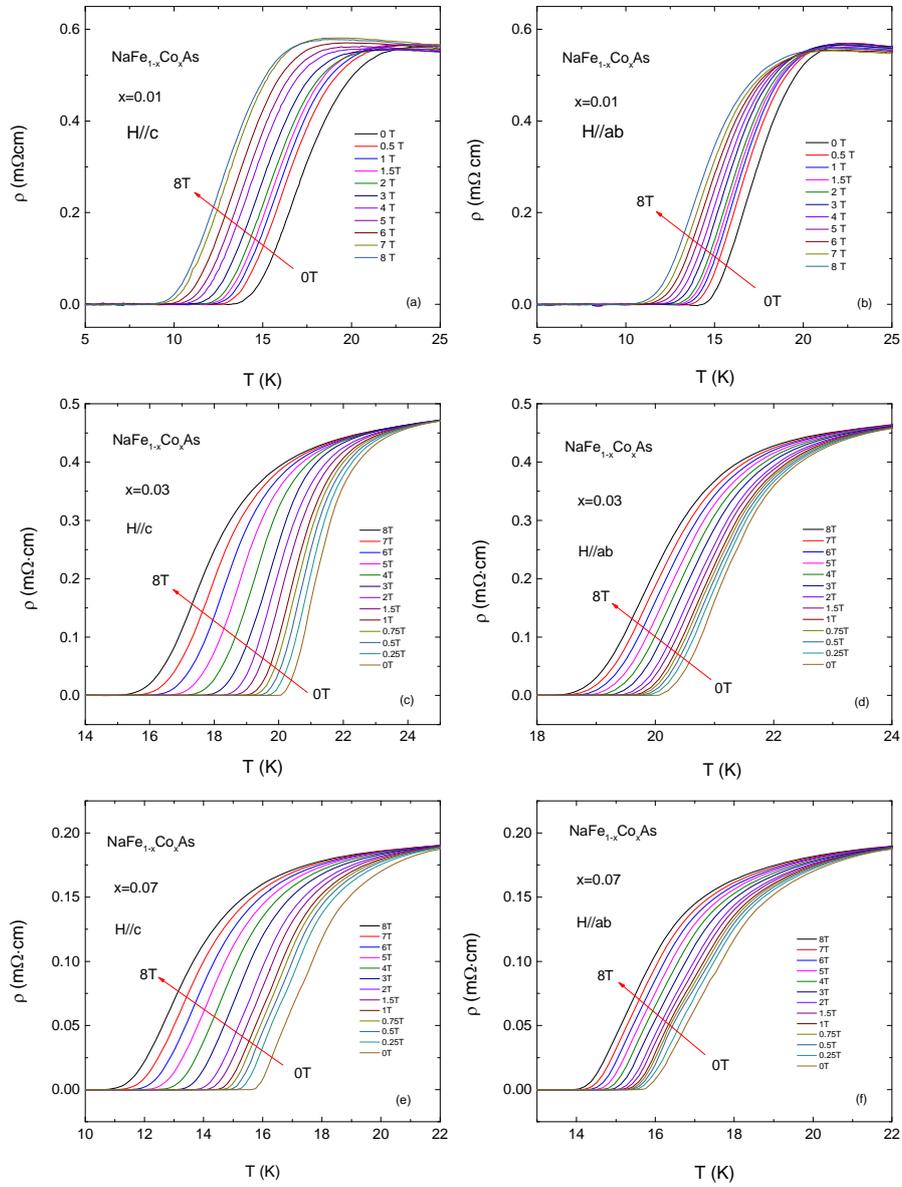

Fig. 4

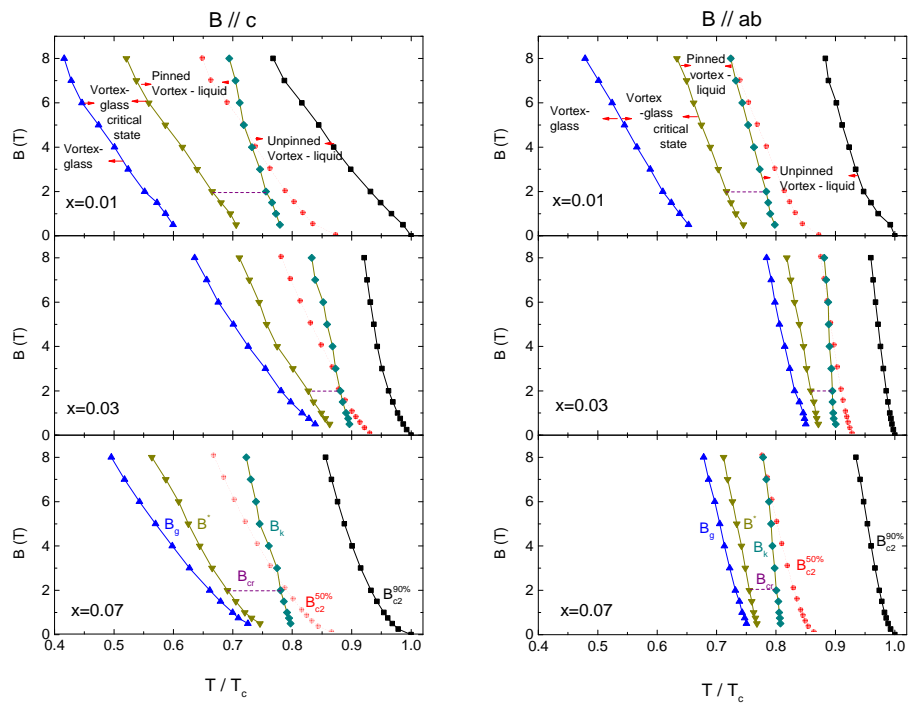

Fig. 5

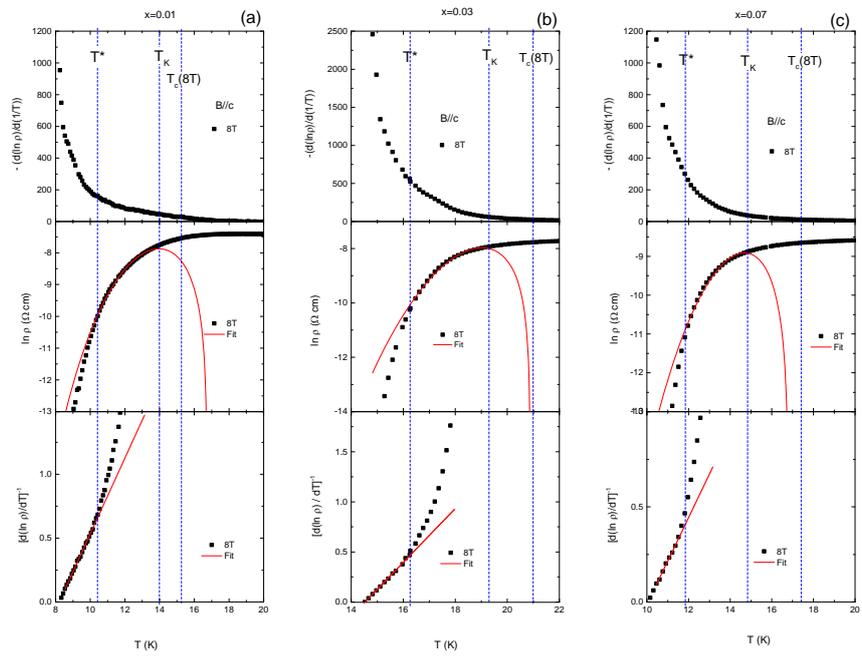

Fig. 6

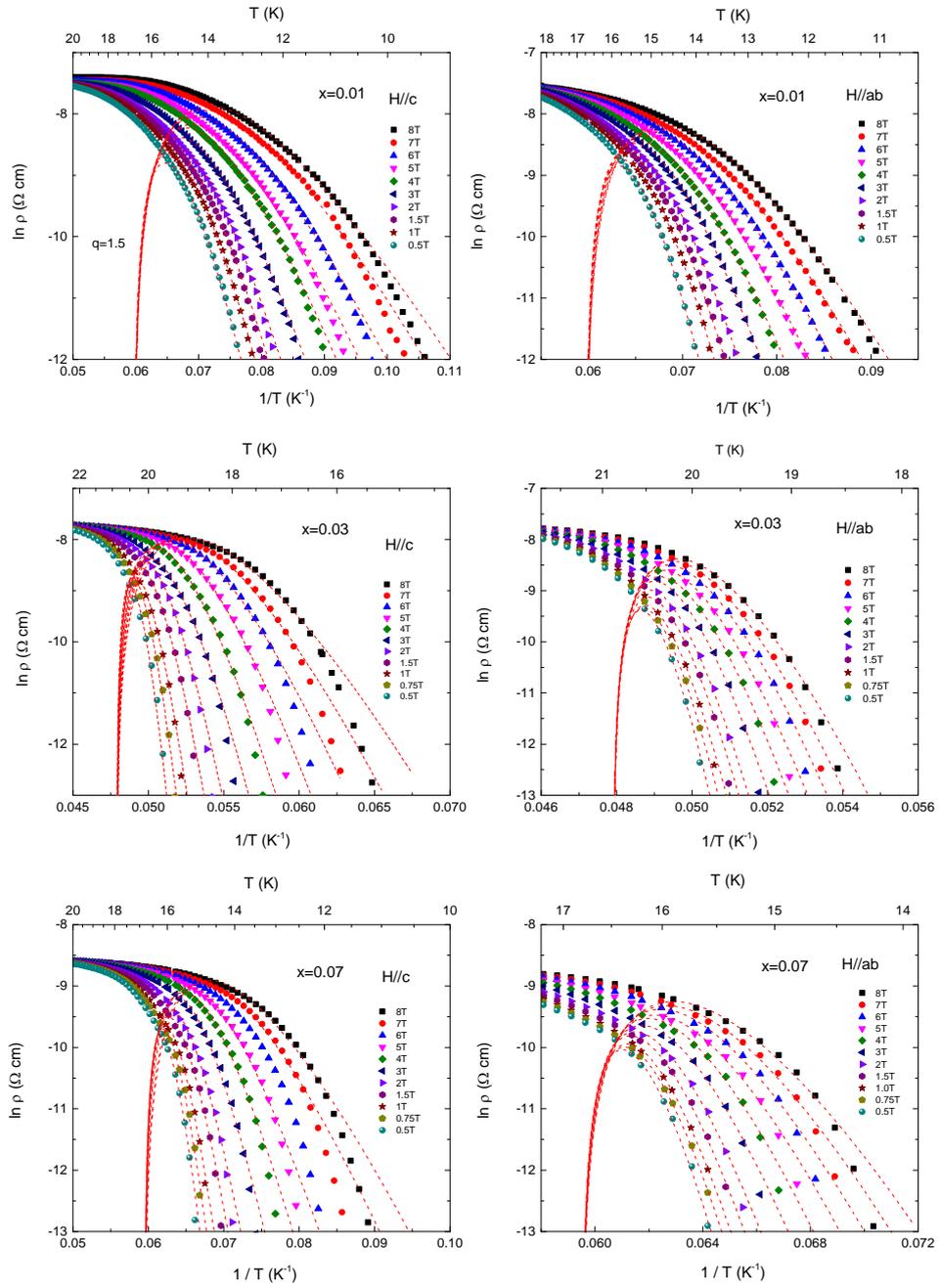

Fig. 7

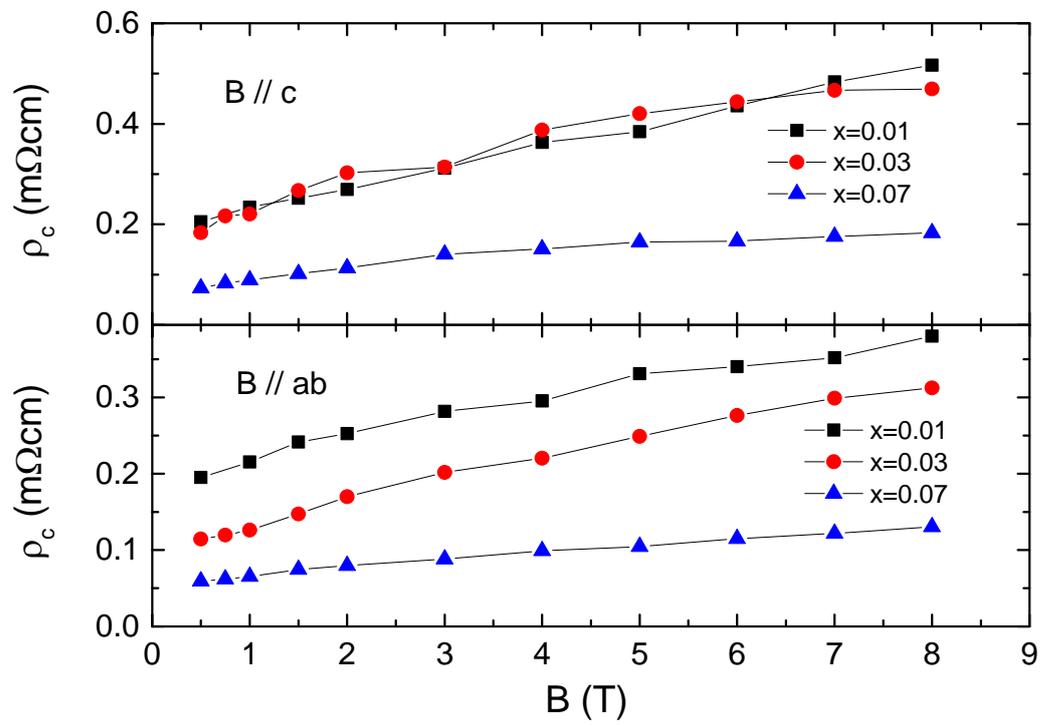

Fig. 8

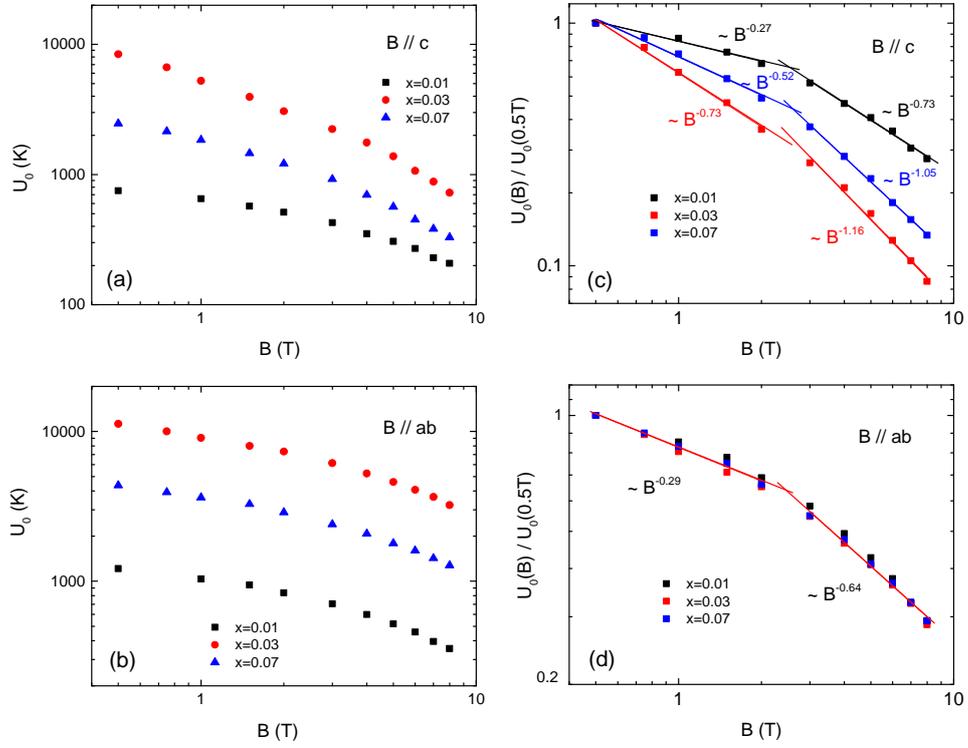

Fig. 9

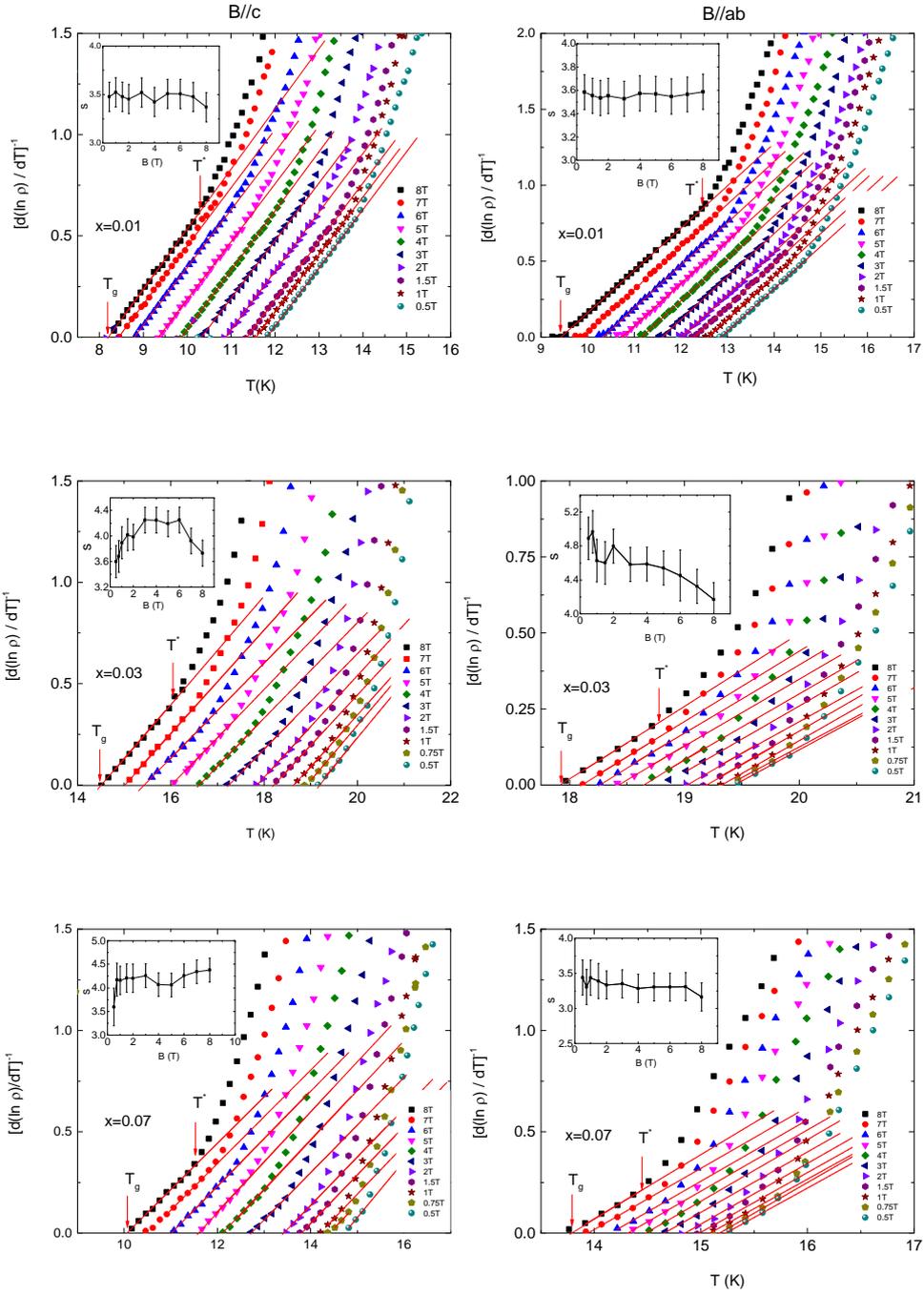